\documentclass[prd,onecolumn,nofootinbib]{revtex4}
\usepackage{bm,amsmath,amssymb,graphicx}

\begin{document}

\noindent
General Relativity and Gravitation {\bf 53} (2), 18 (2021)\\

\title{A nonsingular, anisotropic universe in a black hole\\with torsion and particle production}
\author{Nikodem J. Pop{\l}awski}
\altaffiliation{NPoplawski@newhaven.edu}
\affiliation{Department of Mathematics and Physics, University of New Haven, West Haven, CT, USA}

\begin{abstract}
We consider a universe formed in a black hole in general relativity with spin and torsion.
The interior of a Schwarzschild black hole can be represented by the Kantowski--Sachs metric that describes a closed anisotropic universe.
We use this metric to derive the Einstein--Cartan field equations with a relativistic spin fluid as a source.
We show that torsion may prevent a singularity and replace it with a nonsingular bounce if particle production dominates over shear.
Particle production after the last bounce can generate a finite period of inflation, during which the universe expands and isotropizes to the current state.
This scenario is only approximate: the Kantowski--Sachs metric is never reached and should be replaced with a more general metric that tends to that of a 3-sphere.
\end{abstract}
\maketitle

{\bf 1. Introduction}\\ \\
A recent evidence suggests that the universe may be closed \cite{closed}.
Such a universe might have originated as a baby universe from a bounce in the interior of a black hole existing in a parent universe \cite{Pat,ER,cosmo,ApJ}.
A universe born in a black hole would be located on the other side of the event horizon of the black hole and connected to the parent universe through an Einstein--Rosen bridge \cite{ER}.
Accordingly, its formation and expansion could not be observed outside the black hole because of the infinite redshift at the horizon.
This scenario, which we call Black-Hole Cosmology or Black-Hole Genesis, could naturally solve the black-hole information paradox \cite{cosmo}.
A physical law that turns a black hole into a bridge to a new universe must avoid the black-hole singularity.

A simple and natural mechanism for preventing gravitational singularities is provided by spacetime torsion within the Einstein--Cartan (EC) theory of gravity.
In this theory, expanded by Sciama and Kibble, the Lagrangian density for the gravitational field is proportional to the Ricci scalar, as in the general theory relativity \cite{Lord,EC,Niko}.
EC is therefore the simplest theory of gravity with torsion.
The conservation law for the total (orbital plus spin) angular momentum of fermionic matter in curved spacetime must be consistent with the Dirac equation that allows for the spin--orbit interaction.
This consistency requires that the antisymmetric part of the affine connection: the torsion tensor \cite{Schr}, is not constrained to zero \cite{req}.
Instead, torsion is determined by the field equations obtained from varying the action with respect to the torsion tensor \cite{Lord,EC,Niko}.
The resulting torsion tensor is proportional to the spin tensor of fermionic matter.
Consequently, EC is equivalent to general relativity with the symmetric Levi-Civita connection, in which the energy--momentum tensor of matter acquires terms that are quadratic in the spin tensor.
The multipole expansion of the conservation law for the spin tensor in EC gives a spin tensor which describes fermionic matter as a spin fluid (ideal fluid with spin) \cite{NSH}.

Hehl \cite{Hehl}, Trautman \cite{Tra}, and Kopczy\'{n}ski \cite{Kop1} found that torsion can generate gravitational repulsion and prevent the formation of a cosmological singularity in a homogeneous and isotropic universe described by the Friedmann--Lema\^{i}tre--Robertson--Walker (FLRW) metric \cite{FLRW,Lord,LL2,GR} when spins of fermions are aligned. 
Hehl et al. \cite{HHK} found that macroscopic averaging of the spin terms in the energy--momentum tensor gives a nonzero value even for randomly oriented spins.
Consequently, the effective energy density and pressure of a spin fluid are given by
\begin{equation}
\tilde{\epsilon}=\epsilon-\alpha n_\textrm{f}^2,\quad\tilde{p}=p-\alpha n_\textrm{f}^2,
\label{eff}
\end{equation}
where $\epsilon$ and $p$ are the thermodynamic energy density and pressure, $n_\textrm{f}$ is the number density of fermions, and $\alpha=\kappa(\hbar c)^2/32$ \cite{ApJ,HHK,NP,Gabe}.
At lower densities, the effects of torsion can be neglected and EC effectively reduces to general relativity.
At extremely high densities, much greater than nuclear density, the negative corrections from the spin--torsion coupling in (\ref{eff}) violate the strong energy condition and manifest themselves as repulsive gravity that may prevent the formation of a gravitational singularity in a black hole.
Instead, the collapsing matter undergoes a nonsingular bounce \cite{Nov} and then expands as a new, closed universe \cite{cosmo,ApJ,iso} whose total energy is zero \cite{energy}.

Quantum particle production immediately after the bounce can generate a finite period of exponential inflation \cite{ApJ} that is consistent with the Planck 2015 observations of the cosmic microwave background radiation \cite{SD}.
A bouncing scenario also occurs if the spin tensor is completely antisymmetric \cite{spin}.
Torsion may also explain the matter--antimatter asymmetry in the universe \cite{anti} and the present cosmic acceleration \cite{exp}.
Furthermore, it may eliminate the ultraviolet divergence of loop Feynman diagrams in quantum field theory \cite{toreg}.

The interior of a Schwarzschild black hole can be represented by the Kantowski--Sachs metric (KS) \cite{Ewing}.
This metric describes a closed, anisotropic universe \cite{KS,exact}.
Without further phenomena, such a universe would be singular and oscillatory \cite{osc}, and would not exceed the size of the parent black hole.
If our universe was formed in a black hole, then it must have increased its entropy \cite{ApJ,ent,Garcia} and isotropized in the past to match the currently observed FLRW universe \cite{isotro}.
Both processes can be generated by cosmic inflation \cite{infl}.
Without cosmic acceleration, anisotropy in a closed KS universe would be increasing \cite{pert}.

An anisotropic universe may have past and future singularities that can be avoided under specific conditions \cite{cond,def}.
Kopczy\'{n}ski \cite{Kop2} and Kuchowicz \cite{Kuc1} found that a spatially flat, homogeneous, and anisotropic universe of the Bianchi type I \cite{LL2} has no cosmological singularity if the effect of torsion is greater than that of shear.
Tafel \cite{Taf} found that the avoidance of singularities in EC may also occur for flat, homogeneous universes of the Bianchi types I--VIII.
Further analysis was done in \cite{other} for the Bianchi types with torsion, in \cite{Kuc2} for some particular functions representing the scale factors, and in \cite{Lor} for a spin fluid satisfying the equation of state of stiff matter.

A closed, homogeneous, and anisotropic universe is either of the Bianchi type IX \cite{LL2} or described by the KS metric.
In this article, we consider a KS anisotropic universe, formed in a black hole and filled with a spin fluid.
We investigate its dynamics if the spin fluid is created by quantum particle production \cite{prod}.
After reviewing the interior of a black hole, an anisotropic universe, and the shear tensor, we will demonstrate that such a universe can be nonsingular because of torsion and particle production acting together to violate the strong energy condition.
Furthermore, we will show that its dynamics after the bounce can have a finite period of inflation, during which it produces more matter and isotropizes.
This scenario is only approximate: the exact KS metric is never reached in the presence of torsion and particle production and should be replaced with a more general form that allows the universe to tend to the closed FLRW geometry (a 3-sphere).\\

{\bf 2. Black hole interior metric}\\ \\
The interior of a black hole is given by the Schwarzschild metric:
\begin{equation}
ds^2=-\Bigl(\frac{2GM}{\rho}-1\Bigr)d\tau^2+\Bigl(\frac{2GM}{\rho}-1\Bigr)^{-1}d\rho^2-\rho^2(d\vartheta^2+\sin^2\vartheta\,d\varphi^2),
\label{Schwarz}
\end{equation}
where $M$ is the mass of the black hole, $\tau$ is the external time coordinate, and $\rho<2GM$ is the external radial coordinate \cite{LL2}.
We use units in which $c=1$.
Inside a black hole, the roles of space and time are reversed: $\tau$ becomes an internal spacelike coordinate and $\rho$ becomes an internal timelike coordinate \cite{Ewing}.
Defining a new time coordinate through
\begin{equation}
dt=\Bigl(\frac{2GM}{\rho}-1\Bigr)^{-1/2}d\rho
\label{new}
\end{equation}
and integrating gives a function $t(\rho)$ and then $\rho(t)$.
A new radial coordinate is defined as $r=\tau/b$, where $b$ is a constant.
Substituting $\rho(t)$ into (\ref{Schwarz}) gives \cite{Ewing}
\begin{equation}
ds^2=dt^2-b^2\Bigl(\frac{2GM}{\rho(t)}-1\Bigr)dr^2-\rho^2(t)(d\vartheta^2+\sin^2\vartheta\,d\varphi^2).
\label{interior}
\end{equation}
Defining
\begin{equation}
X(t)=b\Bigl(\frac{2GM}{\rho(t)}-1\Bigr)^{1/2},\quad Y(t)=\rho(t)
\label{XY}
\end{equation}
brings the metric (\ref{interior}) to
\begin{equation}
ds^2=dt^2-X^2(t)dr^2-Y^2(t)(d\vartheta^2+\sin^2\vartheta\,d\varphi^2).
\label{KaSa}
\end{equation}
This is the KS metric \cite{KS}, representing a closed, homogeneous, and anisotropic universe.

The two functions $X$ and $Y$ are related to one another through
\begin{equation}
X=b\Bigl(\frac{2GM}{Y}-1\Bigr)^{1/2}.
\label{cons}
\end{equation}
If the integration in (\ref{new}) uses a condition that $t=0$ at $\rho=2GM$, then the function $t(\rho)$ is \cite{Ewing}
\begin{equation}
t=2GM\Bigl[\Bigl(\frac{\rho}{2GM}\bigl(1-\frac{\rho}{2GM}\bigr)\Bigl)^{1/2}+\arccos\Bigl(\frac{\rho}{2GM}\Bigr)^{1/2}\Bigr].
\end{equation}
The central singularity at $\rho=0$ is reached at $t=\pi GM$.
This instant may be regarded as a cosmological singularity for the universe in a black hole.\\

{\bf 3. Kantowski--Sachs metric}\\ \\
A closed anisotropic universe is described by the interval (\ref{KaSa}), where two scale factors $X(t)$ and $Y(t)$ depend on the cosmic time $t$ and $r,\vartheta,\varphi$ are the spatial coordinates \cite{KS}.
Accordingly, the nonzero components of the metric tensor are $g_{00}=1$, $g_{11}=-X^2$, $g_{22}=-Y^2$, $g_{33}=-Y^2\,\sin^2\vartheta$.
The components of the affine connection are the Christoffel symbols: $\Gamma^{\rho}_{\mu\nu}=\frac{1}{2}g^{\rho\lambda}(g_{\nu\lambda,\mu}+g_{\mu\lambda,\nu}-g_{\mu\nu,\lambda})$, where the comma denotes partial differentiation.
The nonzero components are: $\Gamma^{1}_{01}=\dot{X}/X$, $\Gamma^{2}_{02}=
\Gamma^{3}_{03}=\dot{Y}/Y$, $\Gamma^{0}_{11}=X\dot{X}$,
$\Gamma^{0}_{22}=Y\dot{Y}$,
$\Gamma^{0}_{33}=Y\dot{Y}\sin^2\vartheta$,
$\Gamma^{3}_{23}=\cot\vartheta$,
$\Gamma^{2}_{33}=-\sin\vartheta\,\cos\vartheta$.
Substituting them into the Riemann curvature tensor $R^\lambda_{\phantom{\lambda}\rho\mu\nu}=\Gamma^{\lambda}_{\rho\nu,\mu}-\Gamma^{\lambda}_{\rho\mu,\nu}+\Gamma^{\sigma}_{\rho\nu}\Gamma^{\lambda}_{\sigma\mu}-\Gamma^{\sigma}_{\rho\mu}\Gamma^{\lambda}_{\sigma\nu}$ (we use the notation of \cite{Niko,LL2}) gives $R^1_{\phantom{1}010}=-\ddot{X}/X$, $R^2_{\phantom{2}020}=R^3_{\phantom{3}030}=-\ddot{Y}/Y$, 
$R^2_{\phantom{2}121}=R^3_{\phantom{3}131}=X\dot{X}\dot{Y}/Y$, $R^3_{\phantom{3}232}=\dot{Y}^2+1$.
In the comoving coordinates, $u^0=1$ and the spatial components of the four-velocity vanish.
Accordingly, the nonzero components of the energy--momentum tensor for a spin fluid, $T_{\mu\nu}=(\tilde{\epsilon}+\tilde{p})u_\mu u_\nu-\tilde{p}g_{\mu\nu}$, are $T^0_0=\tilde{\epsilon}$, $T^1_1=T^2_2=T^3_3=-\tilde{p}$.

The Einstein field equations $R^\mu_\nu=\kappa(T^\mu_\nu-1/2\,T\delta^\mu_\nu)$, where $R_{\mu\nu}=R^\rho_{\phantom{\rho}\mu\rho\nu}$ is the Ricci tensor and $\kappa=8\pi G$, are therefore
\begin{eqnarray}
& & R^0_0=-\frac{\ddot{X}}{X}-\frac{2\ddot{Y}}{Y}=\frac{\kappa}{2}(\tilde{\epsilon}+3\tilde{p}), \nonumber \\
& & R^1_1=-\frac{\ddot{X}}{X}-\frac{2\dot{X}\dot{Y}}{XY}=\frac{\kappa}{2}(\tilde{p}-\tilde{\epsilon}), \nonumber \\
& & R^2_2=R^3_3=-\frac{\ddot{Y}}{Y}-\frac{\dot{X}\dot{Y}}{XY}-\frac{\dot{Y}^2}{Y^2}-\frac{1}{Y^2}=\frac{\kappa}{2}(\tilde{p}-\tilde{\epsilon}),
\end{eqnarray}
where the dot denotes differentiation with respect to $t$.
The linear combinations of these equations give \cite{KS}
\begin{eqnarray}
& & \frac{\dot{Y}^2+1}{Y^2}+\frac{2\dot{X}\dot{Y}}{XY}=\kappa\tilde{\epsilon}, \nonumber \\
& & \frac{\dot{Y}^2+1}{Y^2}+\frac{2\ddot{Y}}{Y}=-\kappa\tilde{p}, \nonumber \\
& & \frac{\ddot{X}}{X}+\frac{\ddot{Y}}{Y}+\frac{\dot{X}\dot{Y}}{XY}=-\kappa\tilde{p}.
\label{field}
\end{eqnarray}
The second equation is identical with the Friedmann equation for the scale factor $Y$ and pressure for a closed isotropic universe.
The second and third equation in (\ref{field}) give a relation between $X$ and $Y$ that does not involve the equation of state:
\begin{equation}
\frac{\ddot{X}}{X}-\frac{\ddot{Y}}{Y}+\frac{\dot{X}\dot{Y}}{XY}-\frac{\dot{Y}^2+1}{Y^2}=0.
\label{scale}
\end{equation}

Multiplying the first equation in (\ref{field}) by $XY^2$ and differenting with respect to $t$ gives
\begin{equation}
2\dot{Y}\ddot{Y}X+3\dot{Y}^2\dot{X}+\dot{X}+2\ddot{X}Y\dot{Y}+2\dot{X}Y\ddot{Y}=\kappa\frac{d(\tilde{\epsilon}XY^2)}{dt}.
\label{energy}
\end{equation}
Using the second and third equation in (\ref{field}) gives
\begin{equation}
\kappa\tilde{p}\frac{d(XY^2)}{dt}=-Y^2\dot{X}\Bigl(\frac{\dot{Y}^2+1}{Y^2}+\frac{2\ddot{Y}}{Y}\Bigr)-2XY\dot{Y}\Bigl(\frac{\ddot{X}}{X}+\frac{\ddot{Y}}{Y}+\frac{\dot{X}\dot{Y}}{XY}\Bigr).
\label{pressure}
\end{equation}
Combining (\ref{energy}) and (\ref{pressure}) gives
\begin{equation}
\frac{d(\tilde{\epsilon}XY^2)}{dt}+\tilde{p}\frac{d(XY^2)}{dt}=0.
\label{thermo}
\end{equation}
Equation (\ref{thermo}) is identical with the first law of thermodynamics for an isotropic universe whose scale factor $a$ satisfies $a^3=XY^2$.
We can therefore regard 
\begin{equation}
a=(XY^2)^{1/3}
\label{sf}
\end{equation}
as the volume scale factor, and
\begin{equation}
H=\frac{\dot{a}}{a}=\frac{1}{3XY^2}\frac{d(XY^2)}{dt}=\frac{1}{3}\Bigl(\frac{\dot{X}}{X}+\frac{2\dot{Y}}{Y}\Bigr)
\label{Hubble}
\end{equation}
as the mean Hubble parameter.
Using (\ref{eff}) and (\ref{thermo}) gives the change of entropy:
\begin{equation}
T\frac{d(sXY^2)}{dt}=\frac{d(\epsilon XY^2)}{dt}+p\frac{d(XY^2)}{dt}= \alpha\frac{d(n_\textrm{f}^2 XY^2)}{dt}+\alpha n_\textrm{f}^2\frac{d(XY^2)}{dt},
\end{equation}
where $s$ is the entropy density and $T$ is the temperature.
This relation is similar to that for the FLRW universe \cite{ApJ}.\\

{\bf 4. Bounce}\\ \\
A nonsingular bounce in $Y$ occurs when \cite{def}
\begin{equation}
\frac{\dot{Y}}{Y}=0,\quad \frac{\ddot{Y}}{Y}>0.
\end{equation}
Bounces in $X$ and $a$ are defined similarly.
These definitions are independent of the signs of the scale factors.
If bounces in $X$ and $Y$ exist, then a bounce in $a$ also exists.
These three bounces in general occur at different times.
Volume contraction and expansion of a universe is defined in terms of the sign of $\dot{a}$ (negative for contraction and positive for expansion).\\

{\bf 5. Dust and vacuum solutions}\\ \\
For dust, the pressure of matter is zero.
Consequently, the second equation in (\ref{field}) is $(\dot{Y}^2+1)/Y^2+2\ddot{Y}/Y=0$ and has a parametric solution of a cycloid:
\begin{equation}
t=A\Bigl(\eta+\frac{1}{2}\sin(2\eta)\Bigr),\quad Y=A\cos^2\eta,
\label{parY}
\end{equation}
where $A$ is a constant \cite{KS}.
Equation (\ref{thermo}) gives $\epsilon XY^2=m$, where $m$ is a constant.
Consequently, the first equation in (\ref{field}) leads to
\begin{equation}
X=-B\tan\eta+\kappa m(1+\eta\tan\eta),
\label{parX}
\end{equation}
where $B$ is a constant \cite{KS}.
The rates of $X$ and $Y$ are
\begin{eqnarray}
& & \dot{X}=-\frac{B}{2A\cos^4\eta}+\frac{\kappa m}{2A}\Bigl(\frac{\sin\eta}{\cos^3\eta}+\frac{\eta}{\cos^4\eta}\Bigr), \nonumber \\
& & \dot{Y}=-\tan\eta.
\label{pardot}
\end{eqnarray}
The relations (\ref{parY}), (\ref{parX}), and (\ref{pardot}) determine $a$ and $\dot{a}$ as functions of $\eta$.
For vacuum, the KS metric represents the interior Schwarzschild solution.
In this case, (\ref{parX}) and (\ref{pardot}) give
\begin{equation}
X=B\dot{Y}.
\label{comp}
\end{equation}
This relation also follows from the first and second equation in (\ref{field}) that give $\dot{X}\dot{Y}=X\ddot{Y}$.
Comparing (\ref{comp}) with (\ref{cons}) and using (\ref{new}) and (\ref{XY}) gives $B=b$.
If the integration in (\ref{new}) uses $t=0$ at $\rho=2GM$, then $\eta=0$ at $\rho=2GM$ and (\ref{parY}) gives $A=2GM$.

A KS universe with dust is oscillatory, as a closed universe should be.
The range of the parameter $\eta$ from 0 to $2\pi$ represents one oscillation cycle.
At $\eta=0$, we have $X=\kappa m$, $Y=A$, $\dot{X}=-B/(2A)$, $\dot{Y}=0$, $a=(\kappa mA^2)^{1/3}$, and $\dot{a}=-B/[6A^{1/3}(\kappa m)^{2/3}]$.
In vacuum, this instant corresponds to a disk singularity.
The events $\dot{Y}=0$ and $\dot{a}=0$ do not occur at the same time.
At $\eta=\pi/2$, $X$ diverges, $Y=0$, $\dot{X}$ diverges, $\dot{Y}$ diverges, $a=0$, and $\dot{a}$ diverges, corresponding to a line singularity.\\

{\bf 6. Shear}\\ \\
The shear tensor is the symmetric and traceless part of a tensor $u_{\mu;\rho}(\delta^\rho_\nu-u^\rho u_\nu)$, where $u^\mu$ is the four-velocity and the semicolon denotes covariant differentiation with respect to the Levi-Civita connection \cite{Niko}.
If the four-acceleration vanishes, $w_\mu=u^\nu u_{\mu;\nu}=0$, then the shear tensor is equal to
\begin{equation}
\sigma_{\mu\nu}=u_{(\mu;\nu)}-\frac{1}{3}u^\rho_{\phantom{\rho};\rho}(g_{\mu\nu}-u_\mu u_\nu).
\end{equation}
Using $u_{\mu;\nu}=u_{\mu,\nu}-\Gamma^{\rho}_{\mu\nu}u_\rho$ leads to the nonzero components of this tensor:
\begin{equation}
\sigma^1_1=\frac{2}{3}\Bigl(\frac{\dot{X}}{X}-\frac{\dot{Y}}{Y}\Bigr),\quad \sigma^2_2=\sigma^3_3=\frac{1}{3}\Bigl(\frac{\dot{Y}}{Y}-\frac{\dot{X}}{X}\Bigr).
\end{equation}
The shear scalar is
\begin{equation}
\sigma^2=\frac{1}{2}\sigma^\mu_\nu \sigma^\nu_\mu=\frac{1}{3}\Bigl(\frac{\dot{X}}{X}-\frac{\dot{Y}}{Y}\Bigr)^2.
\label{scalar}
\end{equation}
Accordingly, the shear vanishes and the universe is isotropic if $X$ is proportional to $Y$.
Using the parametric equations for the scale factors gives the shear scalar as a function of $\eta$.
It diverges at $\eta=0$ for vacuum, and at $\eta=\pi/2$.
The expansion scalar is the trace part:
\begin{equation}
\theta=u^\rho_{\phantom{\rho};\rho}=\Gamma^{\rho}_{\rho\nu}u^\nu=\frac{\dot{X}}{X}+\frac{2\dot{Y}}{Y}=3H.
\end{equation}
The rotation tensor $\omega_{\mu\nu}$ is the antisymmetric part of $u_{\mu;\rho}(\delta^\rho_\nu-u^\rho u_\nu)$ and vanishes for any values of $X$ and $Y$.

The rate of the first component of the shear tensor is
\begin{equation}
\dot{\sigma}^1_1=\frac{2}{3}\Bigl(\frac{X\ddot{X}-\dot{X}^2}{X^2}-\frac{Y\ddot{Y}-\dot{Y}^2}{Y^2}\Bigr).
\end{equation}
Using this equation with (\ref{scale}) and (\ref{Hubble}) gives
\begin{equation}
\dot{\sigma}^1_1+3H\sigma^1_1=\frac{2}{3Y^2}.
\end{equation}
Similarly, the rates of the other components satisfy
\begin{equation}
\dot{\sigma}^2_2+3H\sigma^2_2=\dot{\sigma}^3_3+3H\sigma^3_3=-\frac{1}{3Y^2}.
\end{equation}
Consequently, the rate of the shear scalar $d(\sigma^2)/dt=\dot{\sigma}^\mu_\nu\sigma^\nu_\mu$ satisfies
\begin{equation}
\frac{d(\sigma^2)}{dt}+6H\sigma^2=\frac{2}{3Y^2}\Bigl(\frac{\dot{X}}{X}-\frac{\dot{Y}}{Y}\Bigr).
\label{rate}
\end{equation}
The right-hand side in this equation is positive, which indicates that the shear scalar grows with decreasing $a$ faster than $\sim a^{-6}$.\\

{\bf 7. Avoidance of singularity}\\ \\
The Raychaudhuri equation for a congruence of geodesics without four-acceleration and rotation is \cite{Niko}
\begin{equation}
\frac{d\theta}{ds}=-\frac{1}{3}\theta^2-2(\sigma^2-\omega^2)+w^\mu_{\phantom{\mu};\mu}-R_{\mu\nu}u^\mu u^\nu,
\label{Ray}
\end{equation}
where $\omega^2=\omega_{\mu\nu}\omega^{\mu\nu}/2$ is the rotation scalar.
In the KS spacetime in comoving coordinates, $\omega^2=w^\mu_{\phantom{\mu};\mu}=0$.
For a spin fluid, the last term in (\ref{Ray}) is equal to $-\kappa(\tilde{\epsilon}+3\tilde{p})/2$.
Consequently, the necessary and sufficient condition for avoiding a singularity in a black hole is $-\kappa(\tilde{\epsilon}+3\tilde{p})/2>2\sigma^2$.
For a relativistic spin fluid, $p=\epsilon/3$, this condition is equivalent to
\begin{equation}
2\kappa\alpha n_\textrm{f}^2>2\sigma^2+\kappa\epsilon.
\label{avoid}
\end{equation}

Without torsion, the left-hand side of (\ref{avoid}) would be absent and this inequality would not be satisfied, resulting in a singularity.
Consequently, torsion provides a mechanism for preventing a singularity.
However, this mechanism alone is not sufficient.
If the number of fermions in a black hole is constant, then the fermion number density grows with decreasing $a$ according to $\sim a^{-3}$.
Since the shear scalar $\sigma^2$ grows with decreasing $a$ faster than $\sim a^{-6}$, $n_\textrm{f}^2$ growing as $\sim a^{-6}$ could not satisfy (\ref{avoid}) at sufficiently small values of $a$ and a singularity would form.
Therefore, to avoid a singularity, $n_\textrm{f}^2$ must grow faster and fermions must be produced in a black hole.
Quantum particle creation in strongly changing gravitational fields can provide a mechanism that ensures that torsion dominates over shear in the Raychaudhuri equation.
Consequently, torsion and particle production together can replace a singularity in a universe in a black hole with a nonsingular bounce.\\

{\bf 8. Particle production}\\ \\
The interior of a forming black hole would become a vacuum KS universe when all the matter in the black hole reached the central singularity.
However, quantum effects in changing gravitational fields cause particle-pair production that replaces vacuum with a relativistic spin fluid.
These effects may eliminate the singularity and replace it with a nonsingular bounce.

The production rate of particles in a contracting or expanding universe can be phenomenologically given by
\begin{equation}
\frac{1}{\sqrt{-g}}\frac{d(\sqrt{-g}n_\textrm{f})}{dt}=\beta H^4,
\end{equation}
where $g=-X^2Y^4=-a^6$ is the determinant of the metric tensor in (\ref{KaSa}) and $\beta$ is the production rate \cite{ApJ}.
Accordingly,
\begin{equation}
\dot{n}_\textrm{f}+3Hn_\textrm{f}=\beta H^4,\quad \frac{d(n^2_\textrm{f})}{dt}+6Hn^2_\textrm{f}=2\beta n_\textrm{f}H^4.
\label{prod}
\end{equation}
To avoid a singularity, the rate of $n^2_\textrm{f}$ must exceed the rate of $\sigma^2$.
Comparing (\ref{prod}) with (\ref{rate}) gives
\begin{equation}
\frac{2\beta n_\textrm{f}}{81}\Bigl(\frac{\dot{X}}{X}+\frac{2\dot{Y}}{Y}\Bigr)^4>\frac{2}{3Y^2}\Bigl(\frac{\dot{X}}{X}-\frac{\dot{Y}}{Y}\Bigr).
\end{equation}
After the formation of the event horizon, at the instant when (\ref{avoid}) is reached, this inequality must be satisfied to ensure that (\ref{avoid}) continues to hold.\\

{\bf 9. Dynamics of universe with spin fluid}\\ \\
The spin fluid in the early universe is formed by an ultrarelativistic matter in kinetic equilibrium, for which $\epsilon=h_\star T^4$, $p=\epsilon/3$, and $n_\textrm{f}=h_{n\textrm{f}}T^3$, where $T$ is the temperature of the universe, $h_\star=(\pi^2/30)(g_\textrm{b}+(7/8)g_\textrm{f})k_\textrm{B}^4/(\hbar c)^3$, and $h_{n\textrm{f}}=(\zeta(3)/\pi^2)(3/4)g_\textrm{f}k_\textrm{B}^3/(\hbar c)^3$ \cite{ApJ,Gabe}.
For standard-model particles, $g_\textrm{b}=29$ and $g_\textrm{f}=90$.
In the presence of spin and torsion, the first field equation in (\ref{field}) with (\ref{eff}) becomes
\begin{equation}
\frac{\dot{Y}^2+1}{Y^2}+\frac{2\dot{X}\dot{Y}}{XY}=\kappa(h_\star T^4-\alpha h_{n\textrm{f}}^2 T^6).
\label{Niko1}
\end{equation}
Without particle production, (\ref{eff}) and (\ref{thermo}) give the constancy of $XY^2 T^3$, which is equivalent to $\dot{T}/T+H=0$.
With particle production, this constancy is replaced with a relation that follows from the first equation in (\ref{prod}):
\begin{equation}
\frac{\dot{T}}{T}+H=\frac{\beta H^4}{3h_{n\textrm{f}}T^3}.
\label{Niko2}
\end{equation}

Equations (\ref{Niko1}) and (\ref{Niko2}), with (\ref{scale}) and (\ref{Hubble}), determine the time dependence of $X$, $Y$, and $T$.
The only parameters are the mass $M$ of the parent black hole that created the universe and the production rate $\beta$.
The initial conditions at $t=0$ for these equations could be taken as those for the dynamics for the interior of a black hole becoming a KS universe with dust: $X_0=\kappa m$, $Y_0=2GM$, $\dot{X}_0=-B/(4GM)$, and $\dot{Y}_0=0$.
The coordinate $r$ in (\ref{KaSa}) can be rescaled to remove the arbitrariness of the constant $m$.
Consequently, the only arbitrary initial condition is the initial value of $\dot{X}$, related to the constant $B$, which represents the radial velocity of the collapsing matter when the event horizon forms.
If the parent black hole were rotating, then it would be described by the Kerr metric \cite{LL2,Kerr} and the angular momentum of the black hole would be another parameter.
The interior of a Kerr black hole would not be represented by the KS metric; it would be a more general, anisotropic universe.

If $\beta$ is big enough, then particle production will increase $T$ quickly enough to make the right-hand side of (\ref{Niko1}) negative while keeping the left-hand side positive.
This increase will prevent a cosmological singularity and generate a bounce.
After the bounce in $Y$, the universe expands and $Y$ reaches values larger than $2GM$.
Since the universe is closed, $Y$ eventually stops increasing and starts decreasing, and the universe contracts to another bounce.
Because of particle production, which increases entropy, each cycle of expansion and contraction lasts longer and reaches a larger maximum of $Y$ than the preceding cycle \cite{ApJ,ent}.
A similar behavior occurs for $a$.
The universe may have several bounces and cycles until it reaches the size at which the cosmological constant becomes dominant and prevents the next contraction.
Then the cycles end and the universe begins indefinite expansion \cite{ApJ,ent}.
The last bounce would be the big bang.\\

{\bf 10. Inflation and isotropization}\\ \\
Equation (\ref{Niko2}) can be written as
\begin{equation}
\frac{\dot{T}}{T}=H\Bigl(\frac{\beta H^3}{3h_{n\textrm{f}}T^3}-1\Bigr).
\label{Niko3}
\end{equation}
When the mean scale factor $a$ decreases, $H$ is negative and the temperature $T$ increases.
If the production rate $\beta$ is too small, then particle production might be insufficient to prevent a cosmological singularity.
When $a$ increases and if $\beta$ is too big, then the right-hand side of (\ref{Niko3}) could become positive.
In this case, the temperature would grow with increasing $a$, which would lead to eternal inflation \cite{ApJ}.
Consequently, there is a range of the production rate that prevents a singularity and does not cause eternal inflation.

If the right-hand side of (\ref{Niko3}) is slightly lesser than 1, then $T$ would be approximately constant.
Accordingly, $H$ would be also approximately constant and the mean scale factor $a$ would grow almost exponentially, generating inflation.
Since the energy density would be also approximately constant, the universe would produce enormous amounts of matter and entropy.
Such an expansion would last until the right-hand side of (\ref{Niko3}) drops significantly below 1.
Consequently, the period of this inflation would be finite.
After this period, the effects of torsion would become small and the universe would smoothly enter the radiation-dominated expansion, followed by the matter-dominated expansion, and then by the cosmological-constant-dominated expansion.

Inflation in an anisotropic universe leads to its isotropization \cite{infl}.
Equation (\ref{scale}) shows that if $Y\ddot{Y}\gg1$ then $X$ is approximately proportional to $Y$.
Then $a$ is also approximately proportional to $Y$ and satisfies $a\ddot{a}\gg1$, which holds during volume inflation.
Accordingly, the coordinate $r$ in (\ref{KaSa}) can be rescaled to make $X\approx Y$ during this period.
Isotropization would last during inflation, after which it would end and the anisotropy of the universe would grow again during the radiation-dominated era and the matter-dominated era \cite{pert}.
When the universe enters the cosmological-constant-dominated era, the anisotropy would decrease and the closed universe would become asymptotically isotropic.\\

{\bf 11. Summary}\\ \\
The closed geometry \cite{closed} and anisotropy \cite{evi} of the observed universe may indicate that the universe was born in a black hole.
We analyzed a closed Kantowski--Sachs universe satisfying the Einstein--Cartan field equations with a spin fluid.
We showed that torsion and particle production may together violate the strong energy condition and give the universe one or more nonsingular bounces.
Furthermore, particle production in an expanding phase can generate a finite period of inflation, during which the universe produces enormous amounts of matter and entropy, and isotropizes.

This scenario is only approximate.
Its purpose was to show how torsion and particle production could avoid a singularity even if the shear is present.
In reality, the interior Schwarzschild metric and therefore the KS metric, which has topology $R\times S^2$, is never reached if a nonsingular bounce occurs.
It should be replaced with a more general Bianchi IX form, that could allow the universe to asymptotically tend to the closed FLRW geometry with topology $S^3$.

We thank John Barrow for pointing out the role of shear in the early universe.
This work was funded by the University Research Scholar program at the University of New Haven.

\end{document}